\documentclass[cits]{PoS}

\title{Correlated Variability in Blazars}

\ShortTitle{Correlated Variability in Blazars}

\author{\speaker{Robert Wagner}\\
Max-Planck-Institut f\"ur Physik, F\"ohringer Ring 6, D-80805 M\"unchen\\
and\\
Excellence Cluster Universe, Technische Universit\"at M\"unchen, Boltzmannstra\ss e 2, D-85748 Garching\\
        E-mail: \email{robert.wagner@mpp.mpg.de}}

\abstract{Blazars are thought to emit highly-collimated outflows,
so-called jets. By their close alignment to our line of sight,
relativistic beaming effects enable us to observe these jets over
the whole electromagnetic spectrum up to TeV energies, making them
ideal laboratories for studying jet physics. In the last years
multiwavelength observations of blazars provided us with detailed
data sets which helped to characterize the two main components of
the non-relativistic emission, peaking in the optical to X-ray and
GeV/TeV energy region, respectively. In leptonic acceleration
models, they are explained by synchrotron radiation of electrons
and inverse-Compton emission from the same electron population and
thus, correlations of both emission regimes are expected. We
review recent observational results on the presence and absence of
such correlations in blazars, and discuss constraints on emission
models by quantitative correlation analyses.}

\FullConference{Workshop on Blazar Variability across the Electromagnetic Spectrum\\
         April 22-25 2008\\
         Palaiseau, France}

\begin{document}

\section{Blazars}
Blazars are the most extreme class of active galactic nuclei
\cite{r84}. They are thought to have jet structures well-aligned
with our line of sight. Blazars emit their bulk luminosity in the
$\gamma$-ray band and are characterized by high temporal
variability. These short timescales are interpreted by
corresponding emission regions being as small as few light days or
even light minutes only, moving relativistically down the jet.
Blazar spectra are dominated by non-thermal emission, and two
broad, well-defined components can be identified. The low-energy
peak is commonly explained by synchrotron emission from
relativistic electrons in the jet, and observational evidence,
e.g., characteristic emission patterns, but also substantial
polarization, is given. The high-energy peak is explained in
leptonic acceleration models with inverse-Compton (IC) emission
from the same electrons. Particularly for blazars very often
Synchrotron Self-Compton (SSC) \cite{maraschi92} models are
considered (as opposed to external-Compton models,
\cite{dermer92}), which assume the synchrotron photons as the
target field for IC process and the absence of additional strong
contributions by external photon fields. This assumption seems
reasonable for the BL~Lac-type blazars, which often have almost
featureless optical spectra, pointing to a low density of external
photon fields. More generally, of course, IC radiation at
$\gamma$-ray energies can originate from external photons in the
jet, from external clouds, from a dusty torus, or from some other
external seed photon field. Only recently have also low-peaked
BL~Lac objects \cite{magicbllac,wcomae,magics5} and Flat-Spectrum
Radio Quasars (FSRQ) been detected in $E\gtrapprox
75\,\mathrm{GeV}$ $\gamma$-rays \cite{albert3c279}. Particularly
in FSRQ, external photon fields must be considered. Leptonic
scenarios provide a natural explanation for X-ray -- $\gamma$-ray
correlated variability. In hadronic acceleration models, instead,
$\pi^0$ decays from photo-pion production, secondary electrons or
synchrotron emission from protons explain the high-energy peak. In
such acceleration models it is generally more difficult to
accommodate X-ray -- $\gamma$-ray correlations. Also, in hadronic
models simultaneous $\nu$-emission is expected to be observed,
provided sensitive enough $\nu$ detectors \cite{halzenhooper}.

\section{TeV Blazars}
Since 1992, ground-based Imaging Air Cerenkov Telescopes have
proven very successful in detecting high-peaked BL~Lac objects
(HBLs, see, e.g., \cite{wagner08} for an overview, and also
\cite{wystan}). The so-far best-studied blazars are Markarian
(Mkn) 421, Mkn\,501, PKS\,2155--304, and 1ES~1959+650,
particularly because they are known for some years and allowed
extensive studies on various timescales. At the same time, all of
the mentioned objects underwent either dramatic flares (most
prominently Mkn\,501 in early 1997, e.g., \cite{aha99c}, and
recently PKS\,2155--304 \cite{hess2155}) or have rather high duty
cycles (particularly Mkn\,421). Not only are these objects well
studied and well sampled in the TeV range, but also extensive
monitoring and multi-wavelength campaigns have been carried out on
them and are still ongoing.
Note, however, that VHE outbursts do not seem to be a
characteristic property of the TeV blazars in general, and quite
some of the objects discovered since 2002 have not yet shown
significant variability at all \cite{wagner08}.
The TeV HBLs are interesting laboratories for correlation studies
as they offer the possibility of observing the $\gamma$-ray
emission by VHE electrons ($\gamma \approx 10^7$) coupled with
X-ray observations.

\section{SSC Framework}
With generally only weak soft-photon fields in their jet
environment, the high-energy peak of HBLs can be described well by
one-zone or two-zone SSC models. Then, in SSC models, simultaneous
observations of X-rays and $\gamma$-rays are expected to reveal
close correlations of both wavebands, as the corresponding photons
are produced by electrons of similar energies. When, e.g.,
assuming a magnetic field in the acceleration region of $B \approx
0.1\,\mathrm{G}$ and a Doppler factor $\delta \approx10$,
$1\,\mathrm{keV}$ photons are emitted by electrons with
$\gamma\approx10^6$. The very same electrons will up-scatter
photons to $E \sim \gamma m c \sim 1\,\mathrm{TeV}$. Within these
standard parameters, the IC up-scattering takes place in the
Klein-Nishina regime. (To occur in the Thomson regime,
substantially larger values of $\delta$ are required.) Within the
SSC model, simultaneous variability or slight lags for
low-frequencies are expected, accommodated by forward/reverse
shocks \cite{cg99,sokolov04}. A lag of the TeV photons with
respect to the X-ray synchrotron radiation can be explained by the
corresponding IC seed photons (in the optical-UV regime because of
Klein-Nishina suppression of higher-energy photons) becoming
available only after the X-ray synchrotron photons \cite{cg99}. No
such lag is expected in case external seed photons are involved.
Note that also light-travel time effects may lead to time lags in
the SSC scenario.

A linear correlation is generally expected during flare decay,
governed by cooling of the emitting electrons; the seed photons
cool on longer timescale and IC photons should trace the electron
cooling.

Many factors may cause washed-out correlations \cite{marscher04},
particularly if the emission originates from small and localized
acceleration regions, like from shocks \cite{marschergear85}: One
may expect (1) gradients in the highest electron energies, (2) a
superposition (pile-up) of many rising and decaying flares, (3)
that electrons of different energies populate different volumes
(4) that the flare peak is reached earlier at higher frequencies
due to a varying optical thickness.

\section{Observational Caveats}
Correlation between X-ray and $\gamma$-ray emission may provide
essential information on the acceleration mechanism and the
emission properties producing variability. The observational
challenge, however, is to study two light curves obtained
simultaneously by different instruments at largely different
wavebands. Depending on the availability of instruments, the
requirement of simultaneity might not be strictly achievable.
Further on, ideally the sampling rate of the instruments involved
should be similar. Particularly the high-energy emission
($E_\gamma\gtrsim \mathrm{few}\,100\,\mathrm{MeV}$) of blazars, is
characterized by very short variability timescales reaching down
to few minutes in both the X-ray \cite{cat00xue05} and
$\gamma$-ray \cite{albert501,hess2155} regimes. Therefore, not
only an even, but certainly also a dense sampling is mandatory.

\begin{figure}
\includegraphics[width=\linewidth]{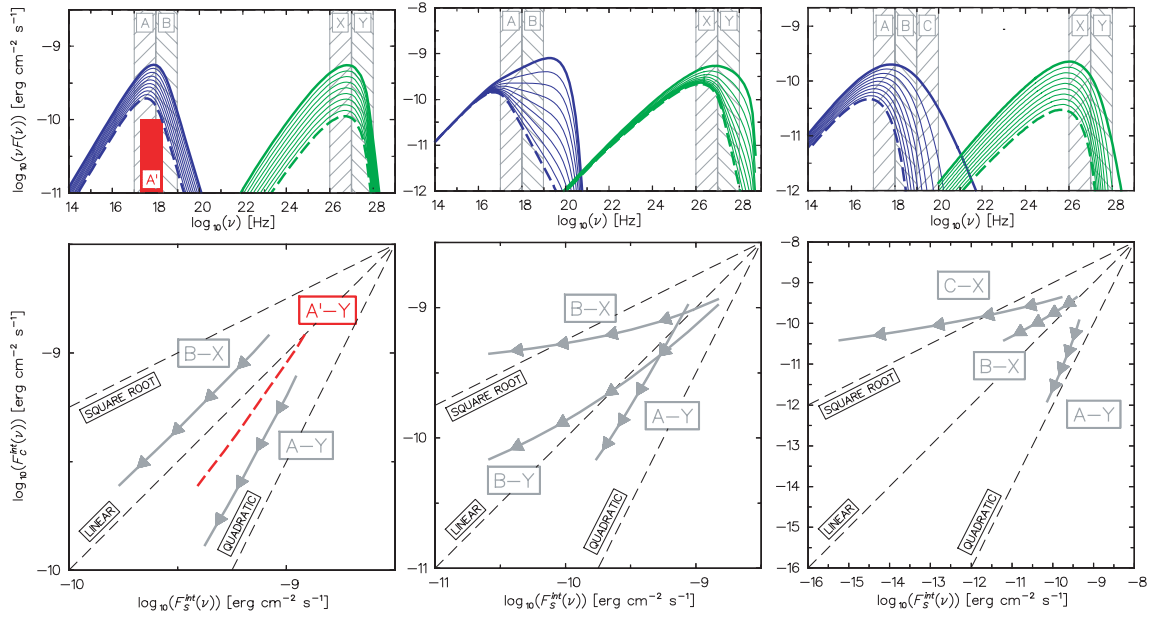}
\caption{Left: Evolution of the SSC emission of an expanding
spherical homogeneous source. The areas A, B, X, Y indicate the
spectral bands used for calculating the correlation. The resulting
correlations are shown in the lower panel. The bold-dashed line
A'--Y shows the correlation obtained after a small shift of the
A--band. Middle: Evolution of the SSC emission of the source where
only the slope $n_2$ of the high-energy part of the electron
spectrum was modified. Same spectral bands A, B, X, Y as in the
left panel. Right: Impact of the radiative cooling. Taken from
\cite{katarzynski05}.} \label{fig:k}
\end{figure}

\paragraph{Systematic studies -- SSC model tests.}
Finally, for any interpretation it is crucial to obtain additional
spectral information to understand the spectral evolution:
Systematic analytical studies \cite{katarzynski05} of the most
simple case of pure SSC scenarios have been carried out assuming a
spherical homogeneous source that may undergo expansion and
compression, changes of the magnetic field, of the electron
density, and adiabatic cooling. Particularly the cases (1) of an
expanding source with constant particle density and magnetic
field, (2) of an increasing particle density while keeping volume
and magnetic field constant, (3) an increasing volume with
adiabatic cooling, and (4) a decrease of particle density while
the source is shrinking, the magnetic field is decreasing and
adiabatic cooling is enabled. Fig.~\ref{fig:k} shows how
correlations are modified by different underlying processes in the
acceleration region, but also how they depend on the position of
the bandpass in both X-ray and $\gamma$-ray regimes at which the
light curves are obtained.

Generally, in SSC models different processes may be considered
responsible for changes in the photon flux \cite{katarzynski05}.
Quadratic correlations can be produced by increasing the electron
density in the source; but these are problematic to sustain during
the decay phase on short timescales. This is because radiative
cooling can only affect high-energy particles. Note also that
strictly a quadratic correlation is only possible for
inverse-Compton processes that take place in the Thomson regime,
in which unacceptably large Doppler factors are implied
\cite{begelman94}. In the Klein-Nishina regime, however,
effectively the low-energy electrons producing the soft seed
photons and the high-energy electrons producing the high-energy
electrons are decoupled. Thus, by providing a constant seed
population of low-energy photons, a linear correlation may be
achieved \cite{katarzynski05}. Rather special and physically
unlikely conditions are needed to construct a quadratic
correlations during decay instead, e.g., adiabatic expansion of
the acceleration region in combination with a constant magnetic
field. Similarly, different source geometries allow for quadratic
decay solutions only with these rather particular physics
conditions. Note that while light-crossing time effects (LCTE) in
the source may weaken intrinsic correlations, substantial external
LCTE may, e.g., modify a quadratic correlation to be observed as a
stronger-than-quadratic one.

\section{Observed Correlations}
Significant evidence of correlated variability has been first
reported for the bright blazars Mkn 421 \cite{buckley96} and Mkn
501 \cite{catanese97}, over observation windows of 15 and 11 days,
respectively. Correlated variability was afterwards observed
during strong flares from Mkn 421, e.g.,
\cite{maraschi99,fossati04,fossati08}, and from Mkn 501
\cite{sambruna00}; over extended periods of weeks to months
\cite{krawczynski00,blazejowski05,gliozzi05,schweizer08}. These
findings were confirmed on other occasions. But correlations were
found to be absent in some cases, most notably in the 2002
``orphan TeV flare'' in 1ES 1959+650 \cite{krawczynski04}; also in
Mkn 421 orphan TeV flares and their counterparts,  ``childless
X-ray flares'' \cite{gliozzi05} were observed.

Comprehensive studies have been carried out by
\cite{krawczynski02} on Mkn 501 by applying a time-dependent SSC
code on a 2-month data set, albeit with a rather sparse sampling;
by \cite{katarzynski05} by systematically taking into account
bandpass effects and changes in different SSC parameters to
account for temporal flux variations as discussed above; and by an
analysis of a week-long, densely-sampled simultaneous data set of
Mkn 421 in 2001 \cite{fossati08}.

\subsection{Correlations on Timescales of Weeks to Months}
\begin{figure}
\begin{minipage}{0.49\linewidth}
\vfill
\includegraphics[width=\linewidth]{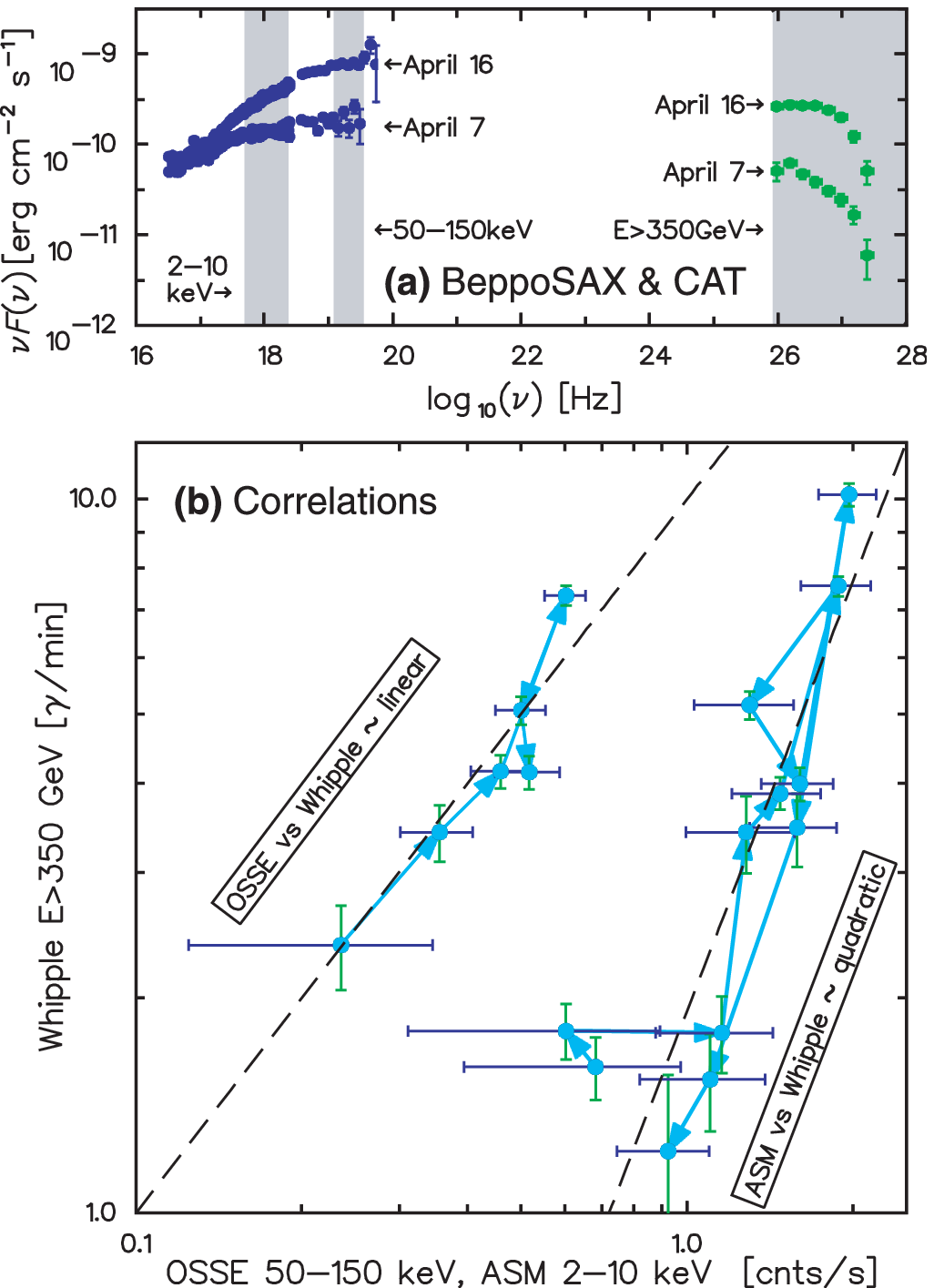}
\caption{The activity of Mkn~501 observed in 1997 April. (a)
Spectra obtained by {\em BeppoSAX} \cite{pian98} and CAT
\cite{djannati99}). The grey areas mark the bandpasses of the
OSSE, ASM, and CAT instruments, respectively. (b) Two correlations
between the X-ray and the TeV $\gamma$-ray fluxes. For clarity,
the OSSE data in (b) were multiplied by 0.5. The dashed lines show
templates for a linear and a quadratic correlation. Taken from
\cite{katarzynski05}.} \label{fig:1}
\end{minipage}
\hfill
\begin{minipage}{0.49\linewidth}
\includegraphics[width=\linewidth]{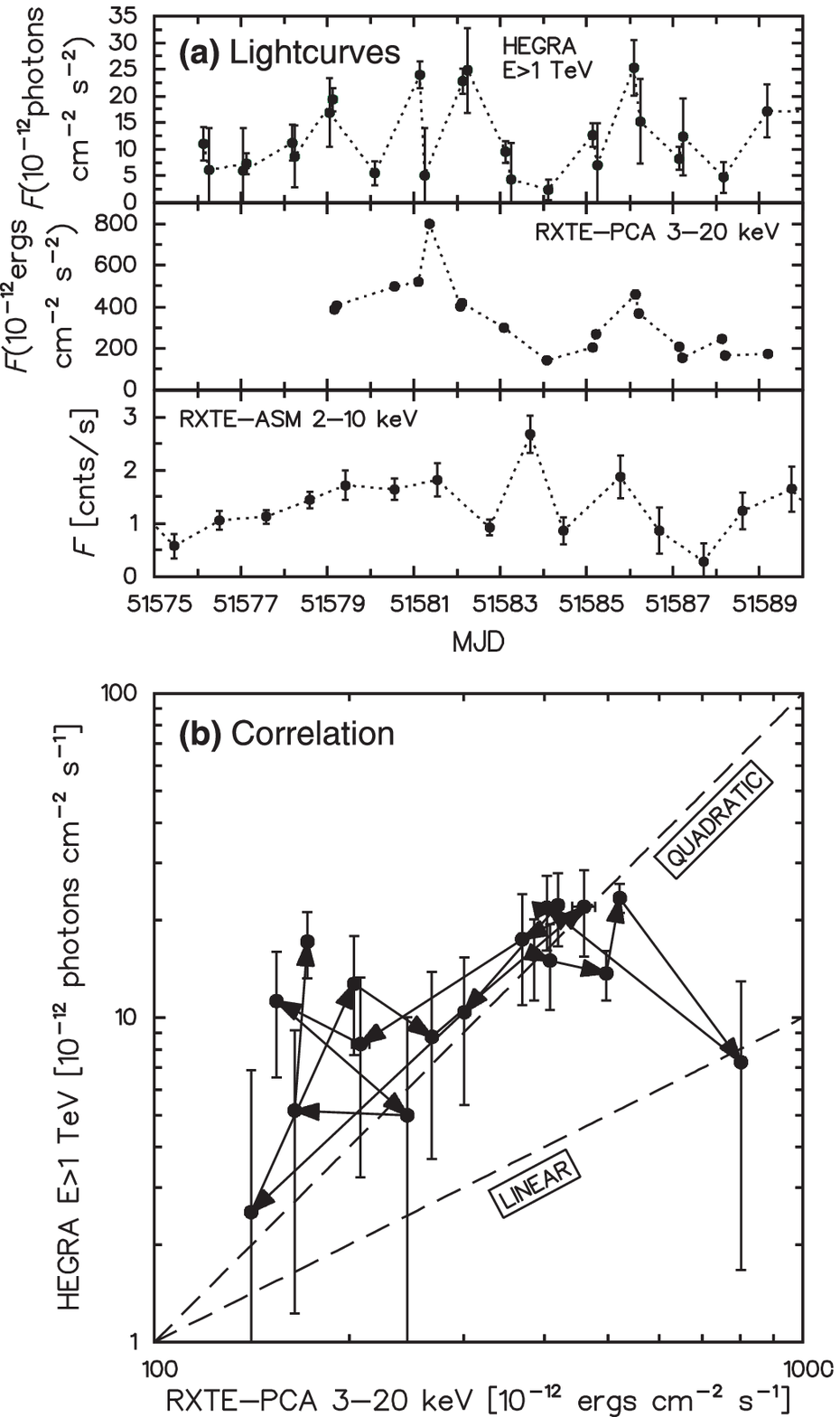}
\caption{The activity of Mkn~421 observed in 2000 February. (a)
$\gamma$-ray and X-ray light curves obtained by HEGRA, {\em
RXTE}-PCA, and {\em RXTE}-ASM \cite{krawczynski01}. (b)
Correlation between the X-ray and the TeV $\gamma$-ray fluxes. The
dashed lines show templates for the linear and the quadratic
correlation. Taken from \cite{katarzynski05}.} \vfill
\label{fig:2}
\end{minipage}
\end{figure}

\paragraph{The Mkn 501 flare in April 1997.}
Mkn 501 underwent a dramatic flare in the first months of 1997,
becoming the brightest $\gamma$-ray source at that time. That
flare was also exceptional as the synchrotron peak energy exceeded
100 keV \cite{catanese97} at peak flare times, and sustained a
peak energy of $\approx$20 keV even in the following year
\cite{pian02}. Observations with the Whipple telescope
($E>300\,\mathrm{GeV}$), with the OSSE instrument
($50-150\,\mathrm{keV}$) on board of {\em CGRO}, and with the
All-Sky Monitor (ASM) instrument aboard the Rossi X-Ray Timing
Explorer ({\em RXTE}; $3-20\,\mathrm{keV}$) yielded correlations
of $F_\gamma \sim F_\mathrm{OSSE}^{1.7 \pm 0.5}$ and $F_\gamma
\sim F_\mathrm{ASM}^{2.7 \pm 0.6}$, respectively
\cite{catanese97}. It was necessary, however, to interpolate the
Whipple and ASM data, as their sampling was less dense than that
of the OSSE data. In this study also correlated optical
variability was found. Observations with CAT and {\em BeppoSAX},
comprising a wider bandpass in the X-rays,
$0.1-200\,\mathrm{keV}$, revealed a similar correlation, $F_\gamma
\sim F_\mathrm{OSSE}^{1.0\dots2.0}$ \cite{djannati99}. Yet another
analysis \cite{krawczynski00} found a quadratic correlation
between the $\gamma$-rays and X-rays measured by ASM
($3-25\,\mathrm{keV}$). While a general correlation between the
two energy regimes was observed, the differences in the
quantitative description of the correlations is due to the
positions and widths of the spectral bands involved (cf.
Fig.~\ref{fig:1}).

\paragraph{The Mkn 421 flare in February 2000.}
Here measurements with the HEGRA imaging air Che\-ren\-kov
telescope array ($E>1\,\mathrm{TeV}$) and the Proportional Counter
Array (PCA) and ASM detectors ($3-20\,\mathrm{keV}$) on board of
{\em RXTE} have been obtained. This time the HEGRA measurements
were interpolated to match the X-ray data. A correlation
($F_\gamma \sim F_\mathrm{X}^{0.5\pm0.2}$) with a rather poor
$\chi^2$ has been found, which makes the interpretation of these
results rather difficult. For this measurement, the rather sparse
sampling of only one data point per day seems to be the limiting
factor.

\subsection{Comprehensive campaigns}
\begin{figure}
\begin{center}
\includegraphics[width=.42\linewidth]{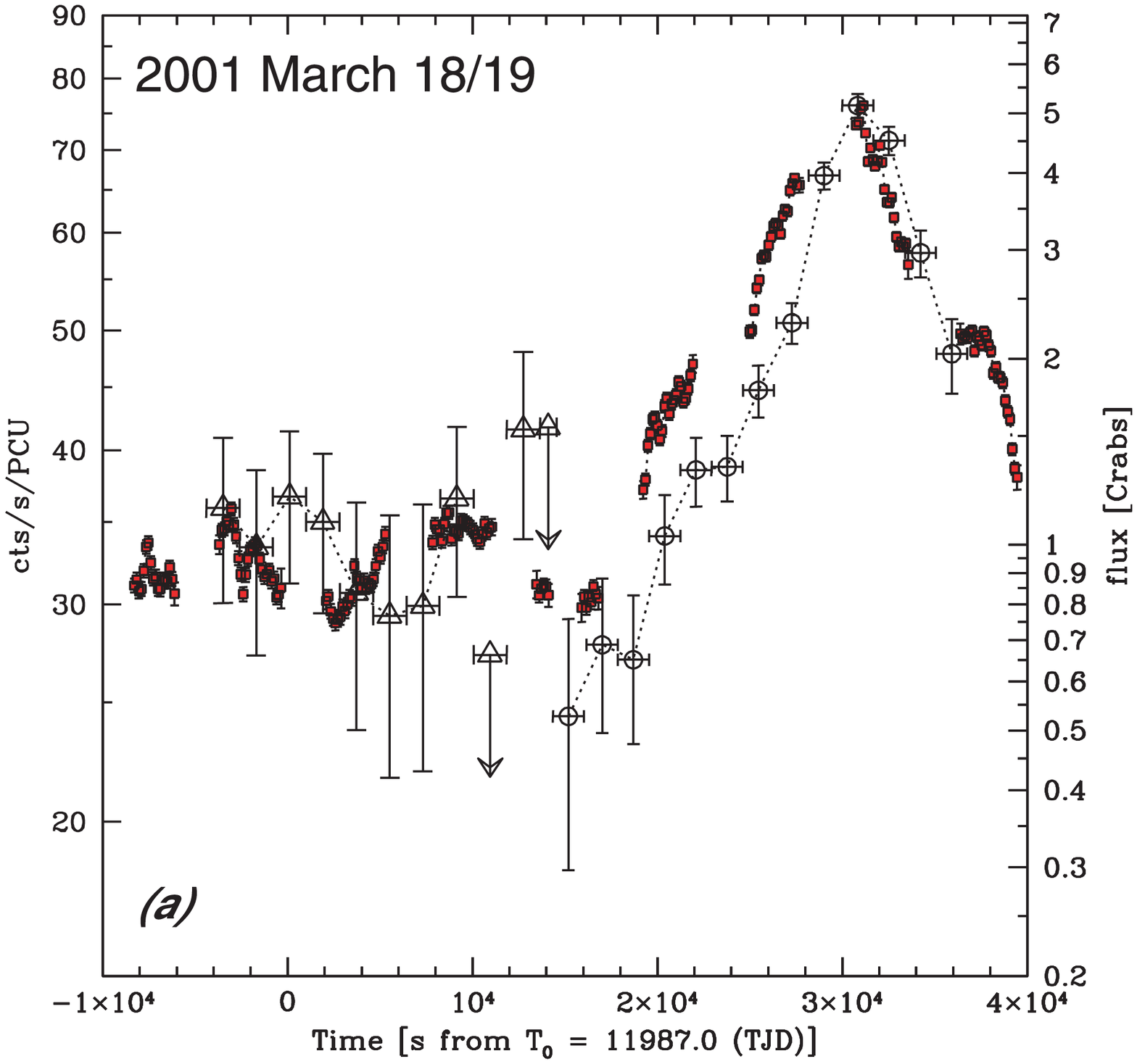}
\includegraphics[width=.42\linewidth]{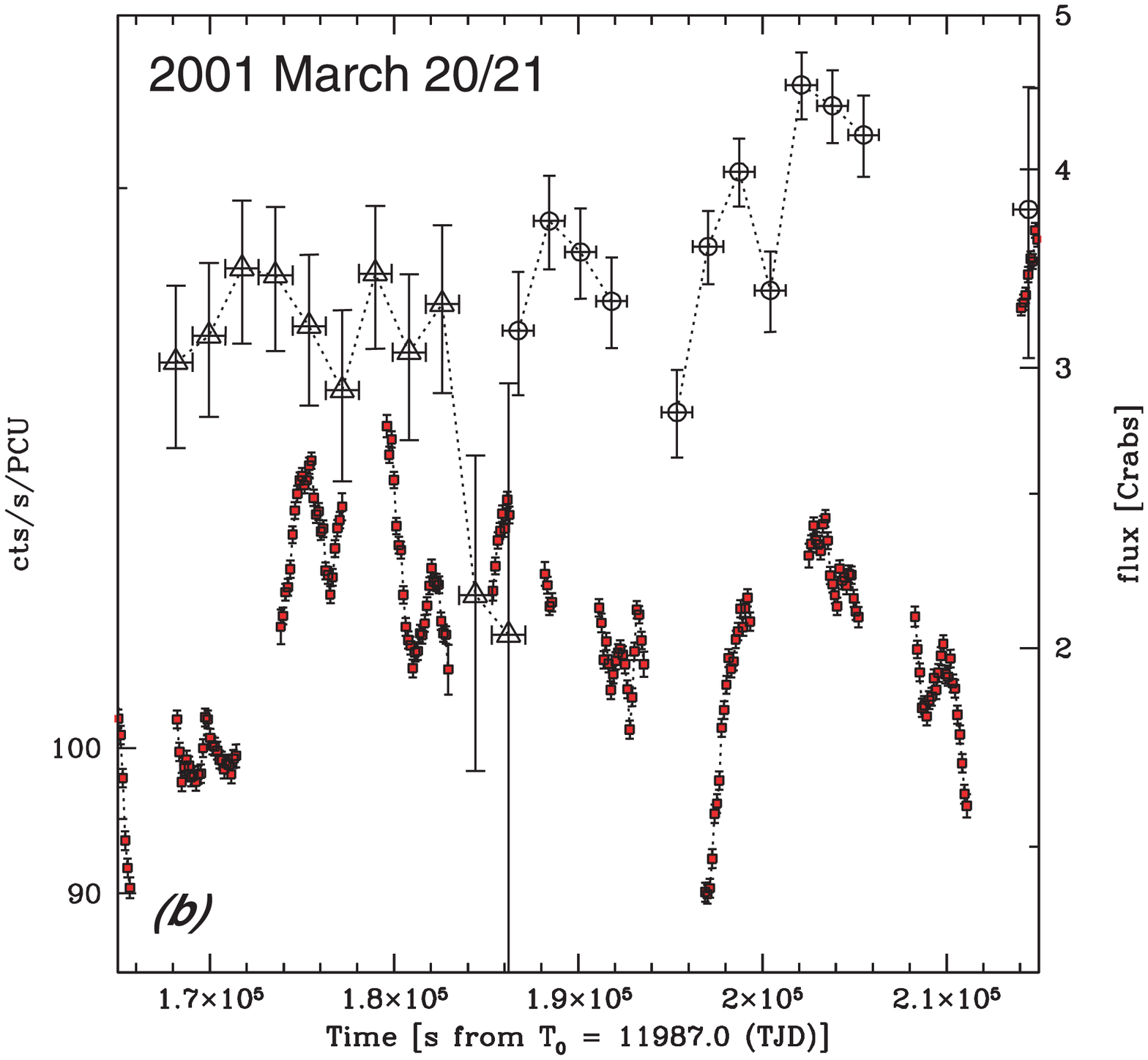}
\end{center}
\caption{Two nights from the 2001 March campaign on Mkn~421. Open
symbols: TeV data (triangles: HEGRA; circles: Whipple). Filled
symbols: PCA data. Strong, highly-correlated variability in both
energy bands is found. Taken from \cite{fossati08}.} \label{fig:3}
\end{figure}

\begin{figure}
\begin{center}
\includegraphics[width=.5\linewidth]{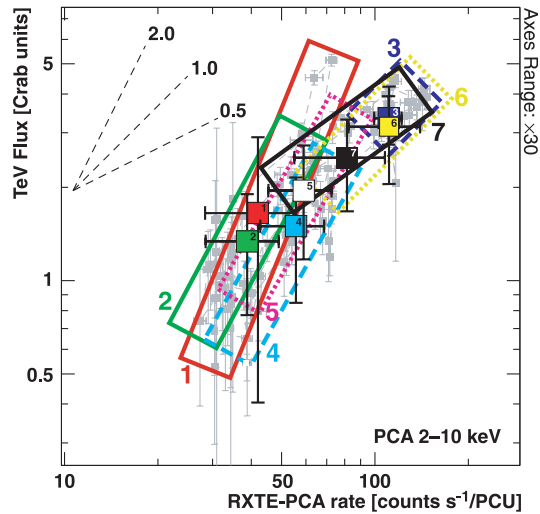}
\end{center}
\vspace{-1cm}
\caption{One-day averaged $\gamma$-ray vs. X-ray
flux. The overall correlation is approximately linear. Numbers
refer to the campaign-night sequence. The boxes approximately
represent the regions of the diagram occupied by the data of each
respective night. The combination of steeper intra-night and
flatter longer-term (due to the shift of the ``barycenters'')
correlations is seen. Taken from \cite{fossati08}.} \label{fig:4}
\end{figure}

\paragraph{The Mkn 421 campaign in  March 2001.}
In 2001 March 19--25, precise observations of Mkn 421, lasting for
one week, were carried out \cite{fossati08}. The good observation
conditions in spring and the combination of data from the Whipple
and HEGRA detectors, some hours apart in longitude (cf., e.g.,
\cite{mazin05}), allowed for a dense sampling in TeV $\gamma$
rays. The PCA instrument aboard {\em RXTE} observed for a total of
62 hours in the $3-20\,\mathrm{keV}$ range during that week.
Generally, clear correlations were observed. Fig.~\ref{fig:3}
shows two of the observation nights. During the night of 2001
March 18/19 (Fig.~\ref{fig:3}a) flare could be for the first time
followed from its onset to its end with sensitive instruments in
both energy bands.
For correlation studies a total of 105 data pairs is available.
Clear correlations between the TeV band and various band passes
ranging from 2 to 60 keV 
in the X-ray regime were found, which generally exhibit linear
correlations. Also the night-by-night averages (7 data points)
follow a linear correlation. The regions, however, which are
covered in the $F_\mathrm{TeV} - F_{\mathrm X}$ plane shift from
night to night, and within these night-by-night observations also
quadratic correlations are found (Fig.~\ref{fig:4}). The fact that
the correlations on day-scale and sub-day scale differ might be
related to more complicated variability patterns, as expected,
e.g., from a pile-up of multiple flares and thus from superimposed
rise and decay phases. An alternative explanation for the complex
picture may also be given by two luminosity-related regimes
between which the emission switches depending on the emission
level.

\begin{figure}
\includegraphics[width=.49\linewidth]{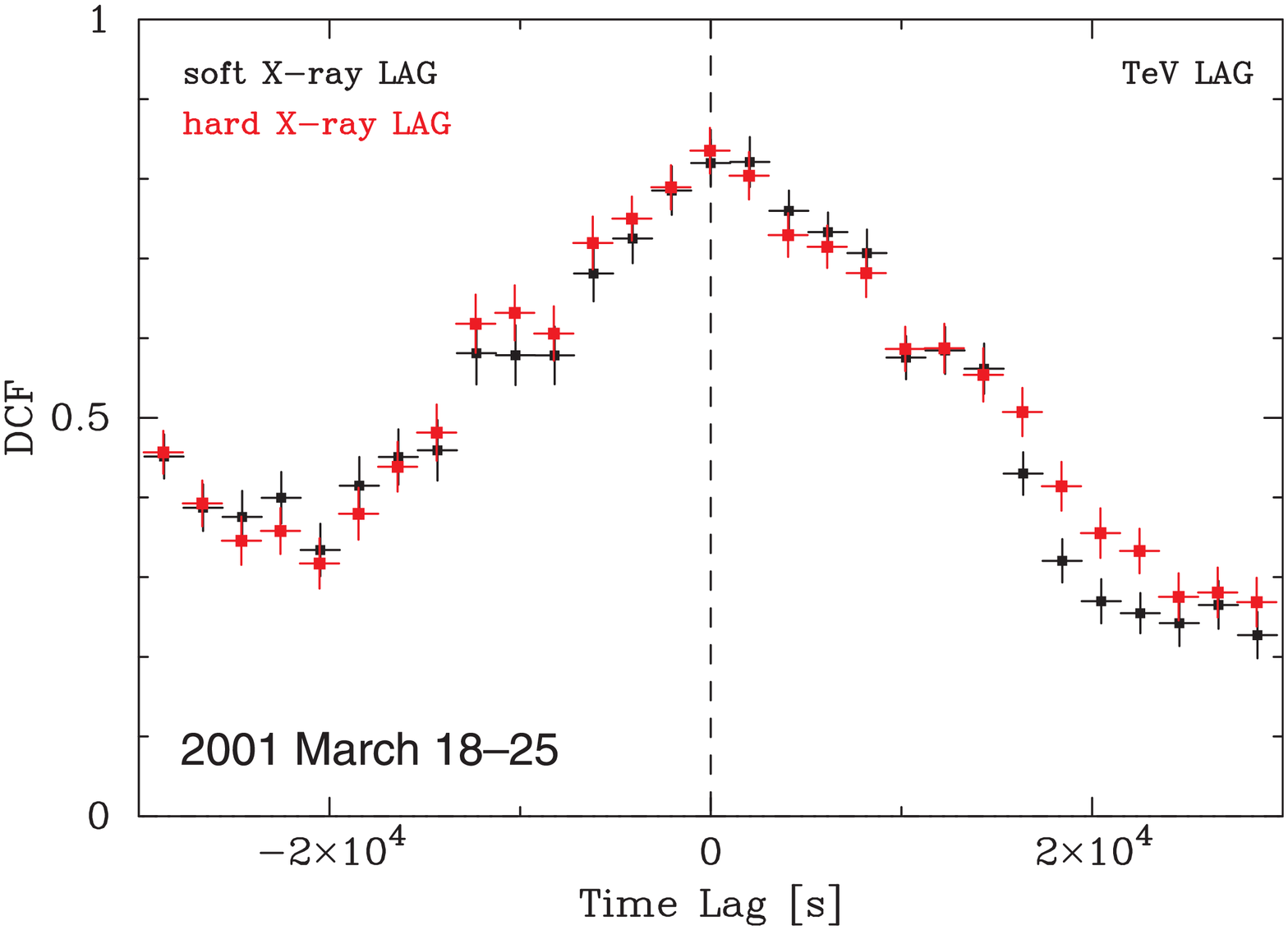}
\includegraphics[width=.49\linewidth]{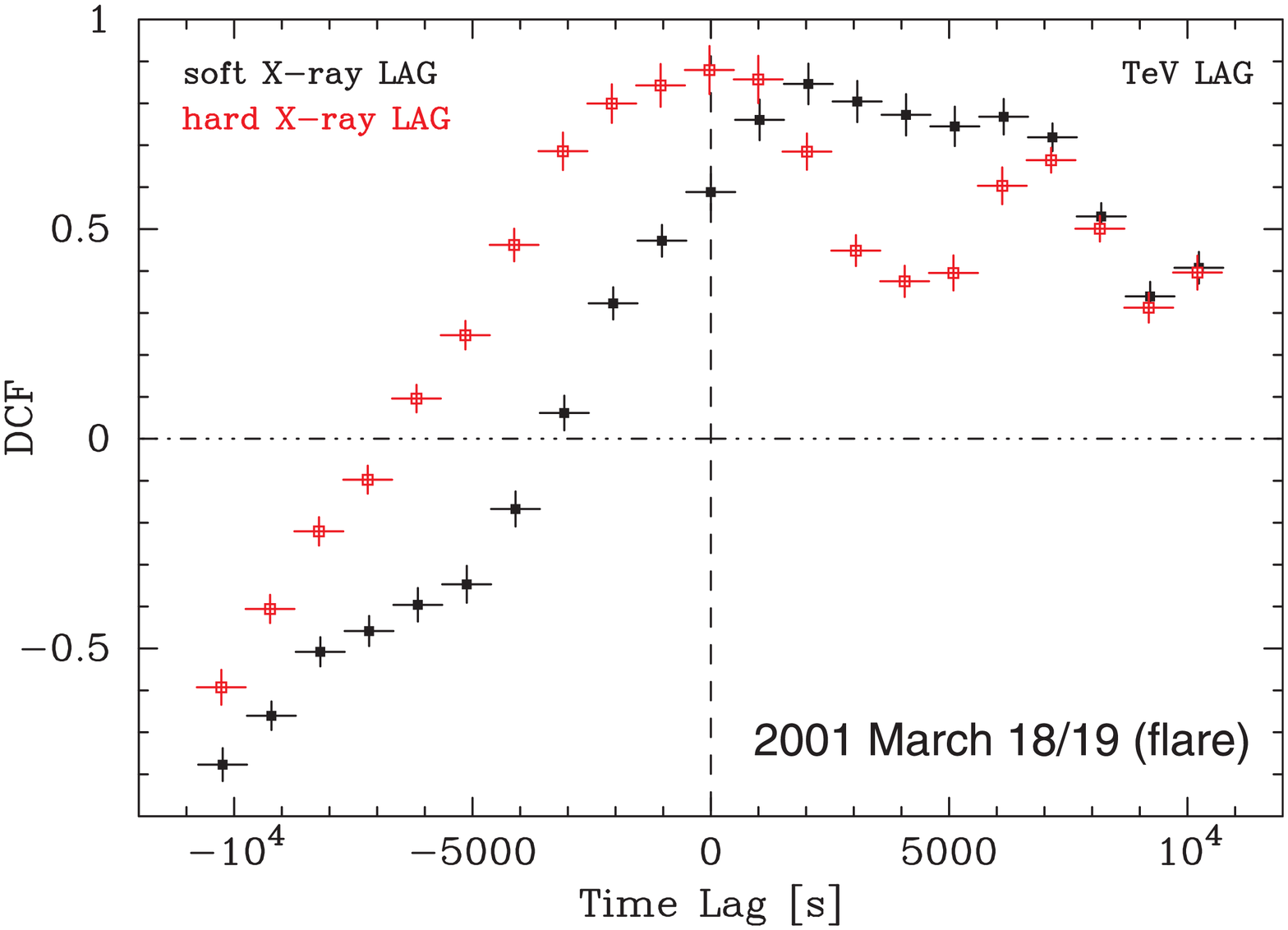}
\caption{Cross-correlation between the X-ray and the TeV light
curves. (a) 2-4~keV vs. TeV $\gamma$-rays (filled, black symbols)
and 9-15~keV vs. TeV (open, red symbols) for the whole campaign.
(b) 2-4~keV vs. TeV (filled, black symbols) and 9-15~keV vs. TeV
$\gamma$-rays (open, red symbols) for the flare night of 2001
March 18-19. Taken from \cite{fossati08}.} \label{fig:5}
\end{figure}

A temporal analysis generally revealed no time lags, while a high
level of correlation, $r=0.8$, was found (Fig \ref{fig:5}a).
Particularly for the March 18/19 flare it is likely that in fact
an undiluted (no unresolved pile-up of many flares) has been
detected. When analyzing the flare night only, a lag of 2~ks for
$\gamma$-rays w.r.t. to soft X-rays is found (Fig.~\ref{fig:5}b),
while the time lag w.r.t. hard X-rays is compatible with zero.

A significantly steeper correlation is found for the flare night
of March 18/19, during which the TeV--X correlation is almost
quadratic during the rising and decay phase of the flare. Note
that particularly during the decay phase a quadratic correlation
is difficult to achieve in SSC scenarios, if the flare decay is
governed by cooling of the emitting electrons. The overall
Whipple--PCA correlation is $F_\gamma \sim
F_\mathrm{X}^{1.3\pm1.0}$. The long-term correlation differs from the
flare correlation. Another 1-week study \cite{tanihata04} of Mkn
421 in 1998 yielded $F_\gamma \sim F_\mathrm{X}^{1.7\pm0.3}$,
where the X-ray flux was measured at the synchrotron peak
position.

\subsection{The Orphan Flare Case}
\begin{figure}
\begin{center}
\begin{minipage}{0.66\linewidth}
\vfill
\includegraphics[width=.99\linewidth]{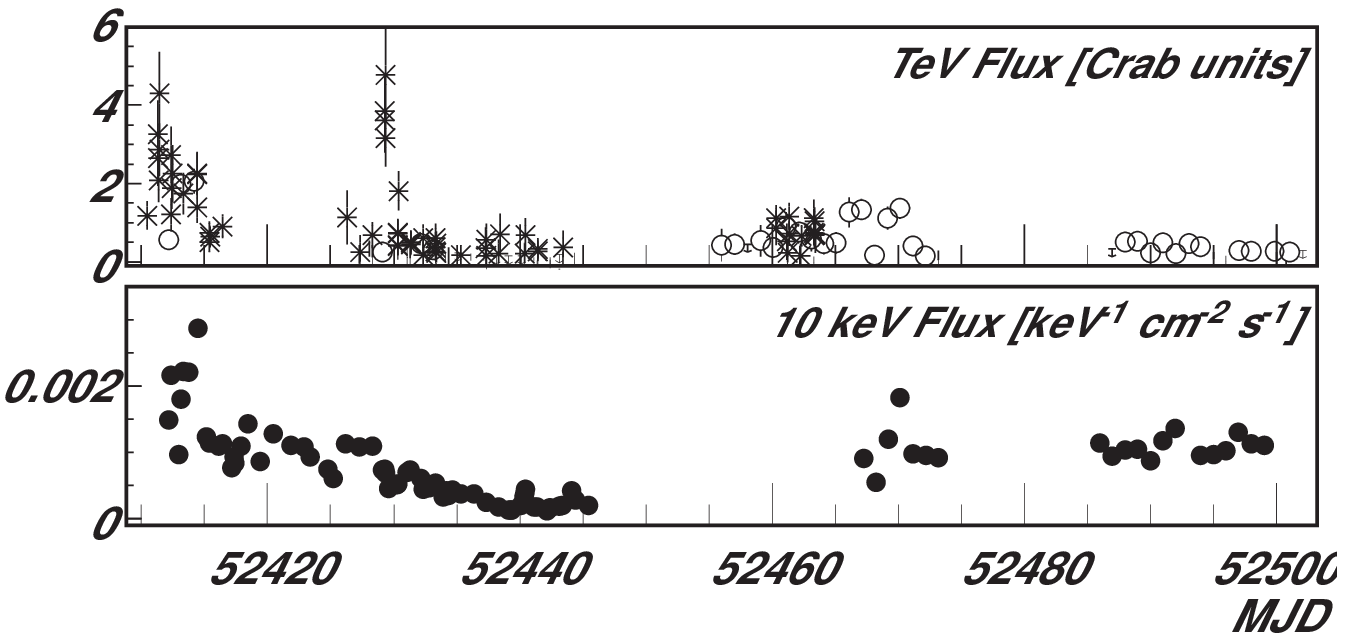}
\caption{Results from the 1ES\,1959+650 multiwavelength campaign
(2002 May 16 -- August 14), taken from \cite{krawczynski04}. Top:
Whipple (stars) and HEGRA (circles) integral TeV $\gamma$-ray
fluxes; (bottom) {\it RXTE} X-ray flux at 10 keV.} \label{fig:es1}
\vfill
\end{minipage}
\hfill
\begin{minipage}{0.32\linewidth}
\includegraphics[width=.99\linewidth]{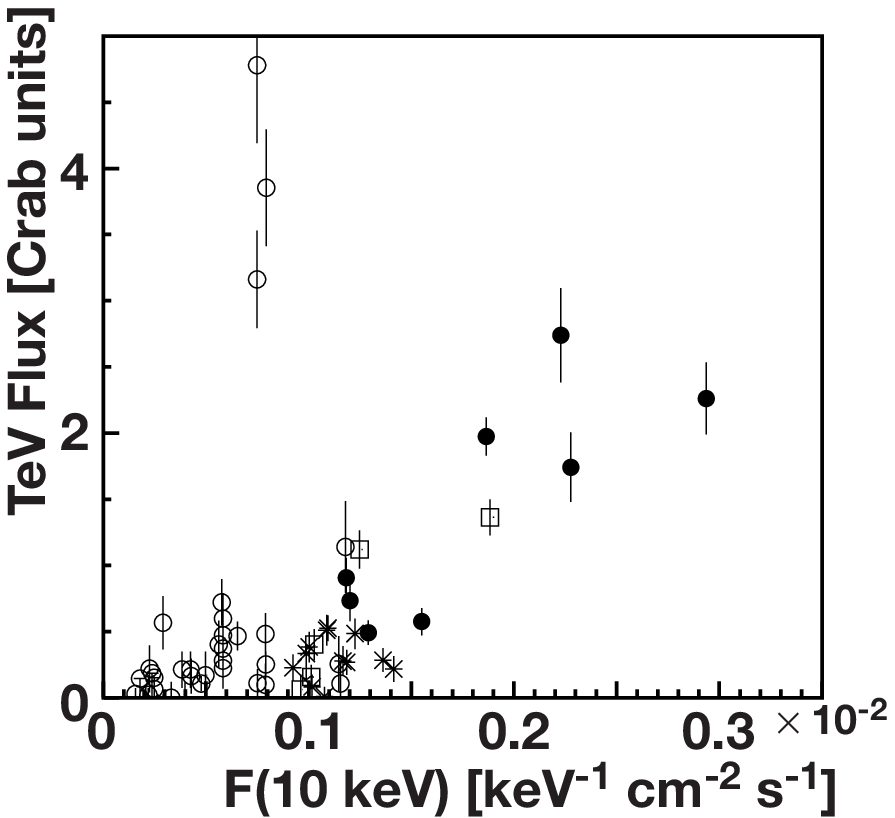}
\caption{Correlation between the X-ray flux and the 3-25 keV
photon index. The different symbols represent different epochs of
the campaign. Taken from \cite{krawczynski04}.} \vfill
\label{fig:es2}
\end{minipage}
\end{center}
\end{figure}

In 2002, strong flares from 1ES 1959+650 were observed
\cite{krawczynski04} (Fig.~\ref{fig:es1}). Optical, X-rays ({\em
RXTE}-PCA), and $\gamma$-rays (Whipple) in general showed a good
correlations. Particularly also a tight correlation between the
3-25 keV spectral index and the 10 keV flux was found
(Fig.~\ref{fig:es2}), pointing to a synchrotron peak moving
towards higher peak energies with increasing flux. However, the
data set also shows clear evidence for an ``orphan flare'', i.e. a
TeV flare without any obvious counterpart in the X-ray band. Such
flares can be explained in hadronic acceleration models, and it
was noted that there was a preceding X-ray flare about 15 days
before the TeV flare, with the TeV flare possibly being connected
to that earlier event. A couple of emission models have been
discussed in the literature
\cite{krawczynski04,boettcher05,reimer05}, one of them invoking
broad-line region clouds as mirrors for the primary flare, and
subsequent secondary flare from $\mathrm{p}\gamma$-interactions.

\subsection{Long-term Studies: Timescales of Years}
\begin{figure}
\begin{center}
\includegraphics[width=\linewidth]{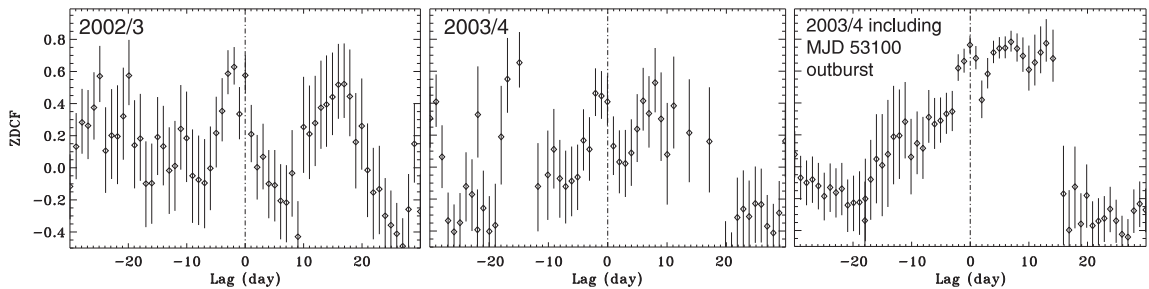}
\caption{Z-transformed discrete correlation function between the
X-ray and TeV light curves for Mkn 421 in the 2002/3 season
(left), in the 2003/4 season before the giant outburst on MJD
53100 (middle), and in the entire 2003/4 season. Taken from
\cite{blazejowski05}.} \label{fig:6}
\end{center}
\end{figure}

\begin{figure}
\begin{center}
\includegraphics[width=.70\linewidth]{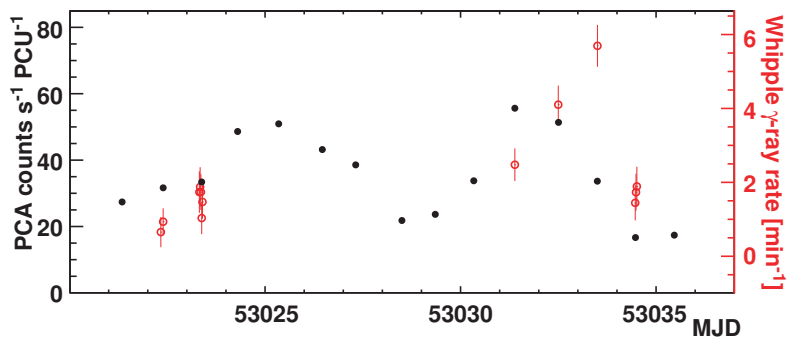}
\caption{A TeV (red, open circles) flare in Mkn\,421 without
simultaneous X-ray counterpart. The (black, filled dots) X-ray PCA
light curve underwent significant variability and peaked $\approx
1.5$ days before the TeV flare. Taken from \cite{blazejowski05}.}
\label{fig:7}
\end{center}
\end{figure}

\paragraph{Mkn 421 in the 2003/4 Observing Season.}
The Whipple telescope observed Mkn 421 regularly during dark time
in the 2003/4 season, for about 28 minutes per observation, with
more runs on occasion \cite{blazejowski05}. In an analysis of 306
such runs a rough correlation was found with {\em RXTE}-PCA X-ray
data. As both the Whipple and PCA energy bands are rather broad,
this weak correlation likely is not an observational artifact.
Despite an acceptable dynamical range 30 in both energy bands and
the broad spectral coverage, the correlation more loose than
expected. The discrete correlation function shows a marginal lead
of the X-ray w.r.t. the $\gamma$-ray light curve of $1.4\pm0.8$
days (Fig.~\ref{fig:6}) both in the 2002/3 observing season and in
the 2003/4 season before an outburst on MJD 53100. When including
this flare, however, the picture is modified by the different
X-ray and $\gamma$-ray rise times of the flare. The flare itself
can be categorized as ``orphan TeV flare'' without simultaneous
counterpart, where a candidate X-ray counterpart may have peaked
1.5 days earlier (Fig.~\ref{fig:7}).

\begin{figure}
\centering
\includegraphics[bb=35 30 355 296,clip=,angle=0,width=.33\linewidth]{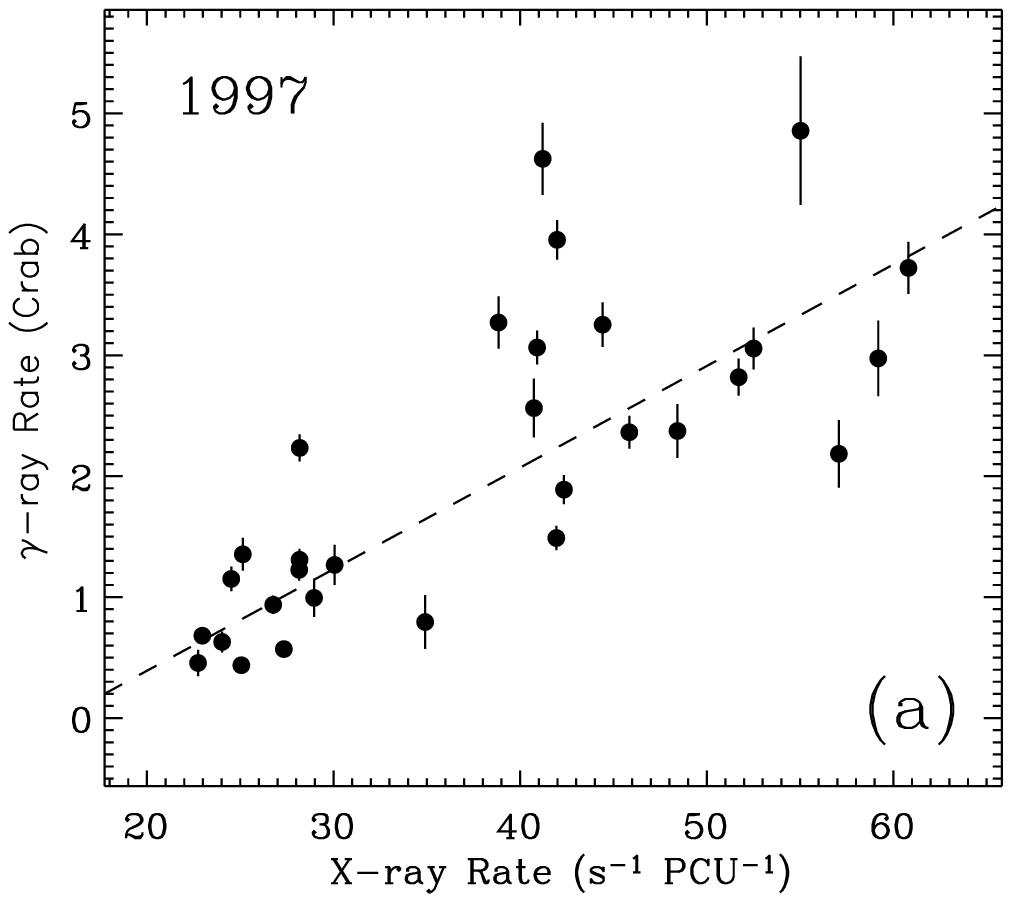}\includegraphics[bb=35 30 355 296,clip=,angle=0,width=.33\linewidth]{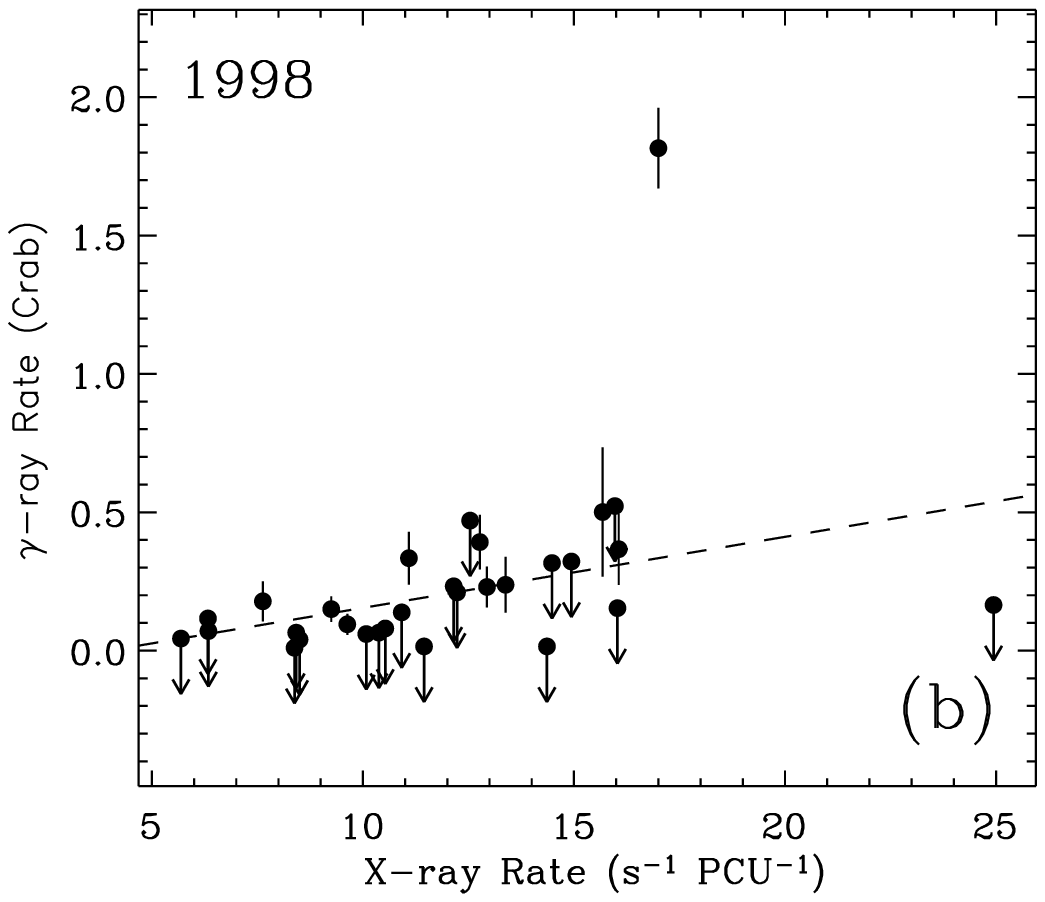}
\includegraphics[bb=35 30 355 296,clip=,angle=0,width=.33\linewidth]{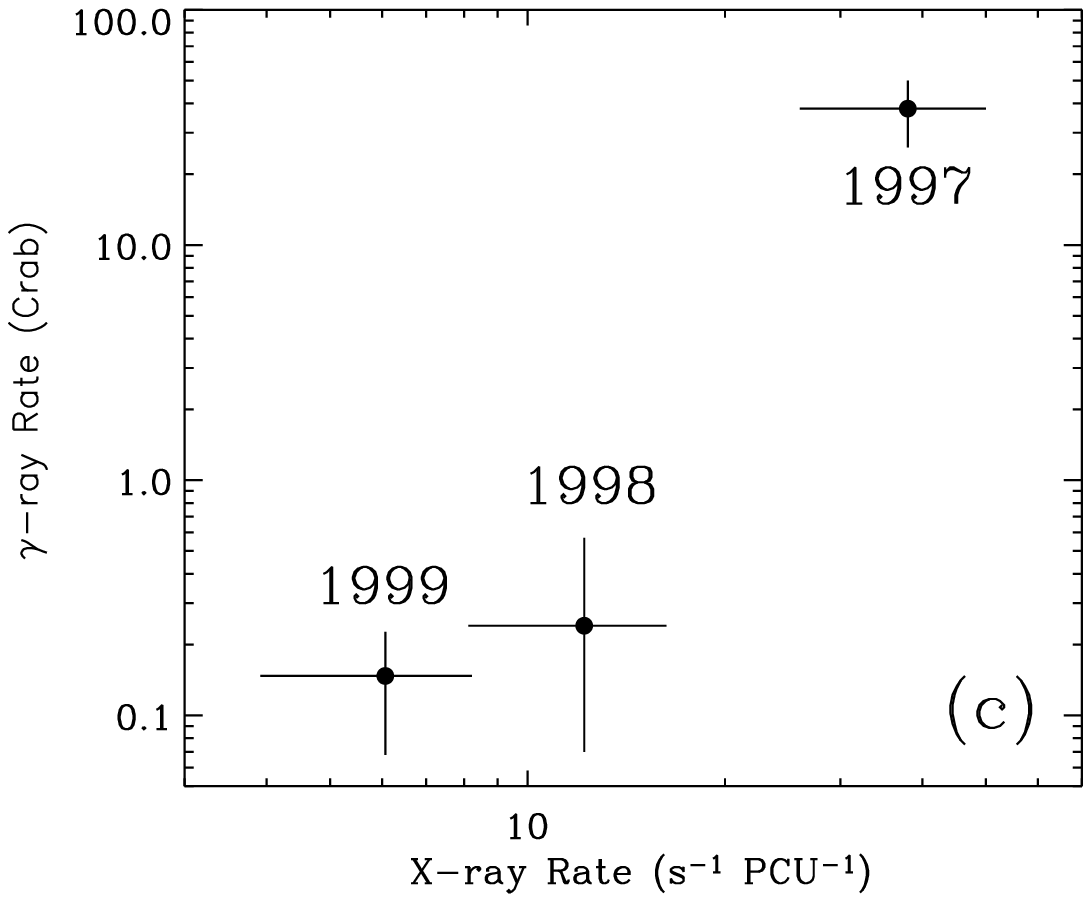}
\caption{Mkn\,501 $\gamma$-ray vs. the 2-20 keV X-ray count rate
during (a) 1997 and (b) 1998, respectively. (c) Mean values of the
$\gamma$-ray fluxes (calculated using all data sets) vs. X-ray
count rates; the horizontal and vertical error bars represent the
dispersion around the X-ray and $\gamma$-ray mean values. Taken
from \cite{gliozzi05}.} \label{fig:8}
\end{figure}

\begin{figure}
\centering
\includegraphics[bb=55 60 295 320,clip=,angle=0,width=.29\linewidth]{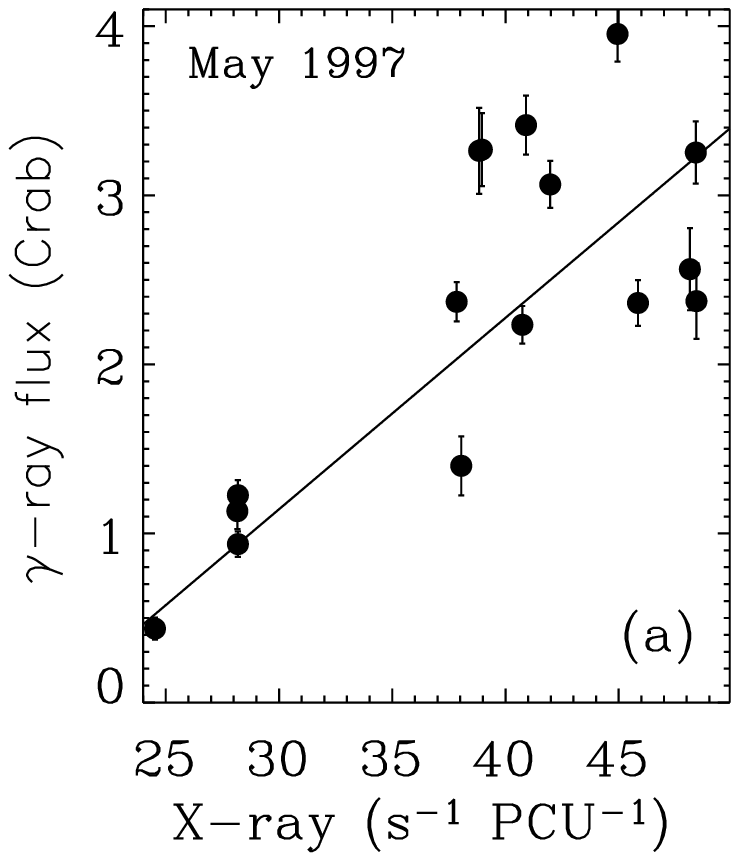}\includegraphics[bb=55 60 295 320,clip=,angle=0,width=.29\linewidth]{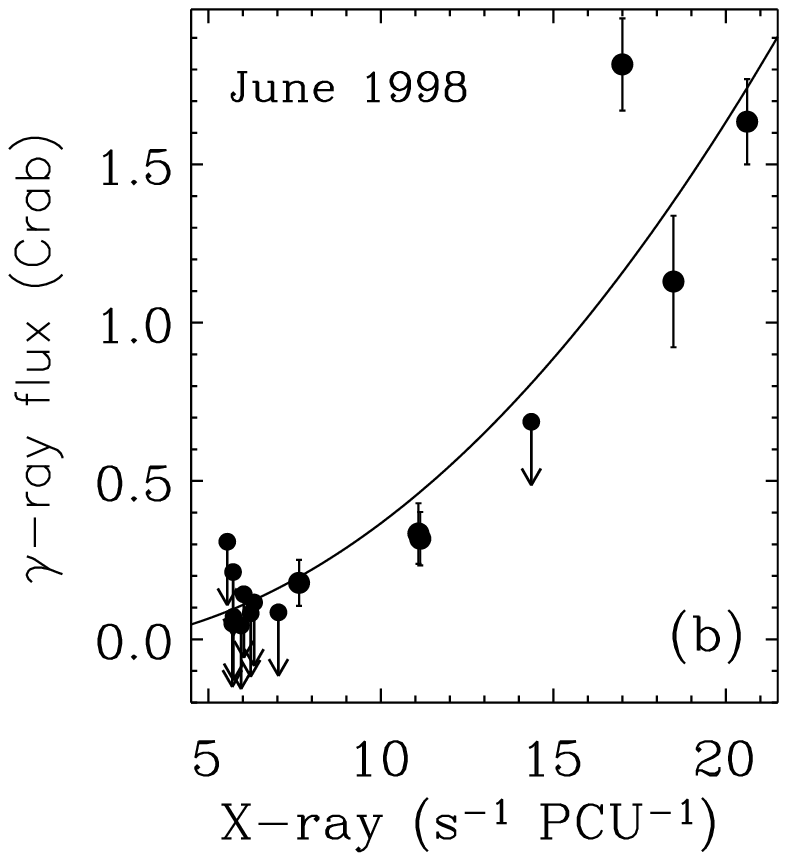}
\includegraphics[bb=55 60 295 320,clip=,angle=0,width=.29\linewidth]{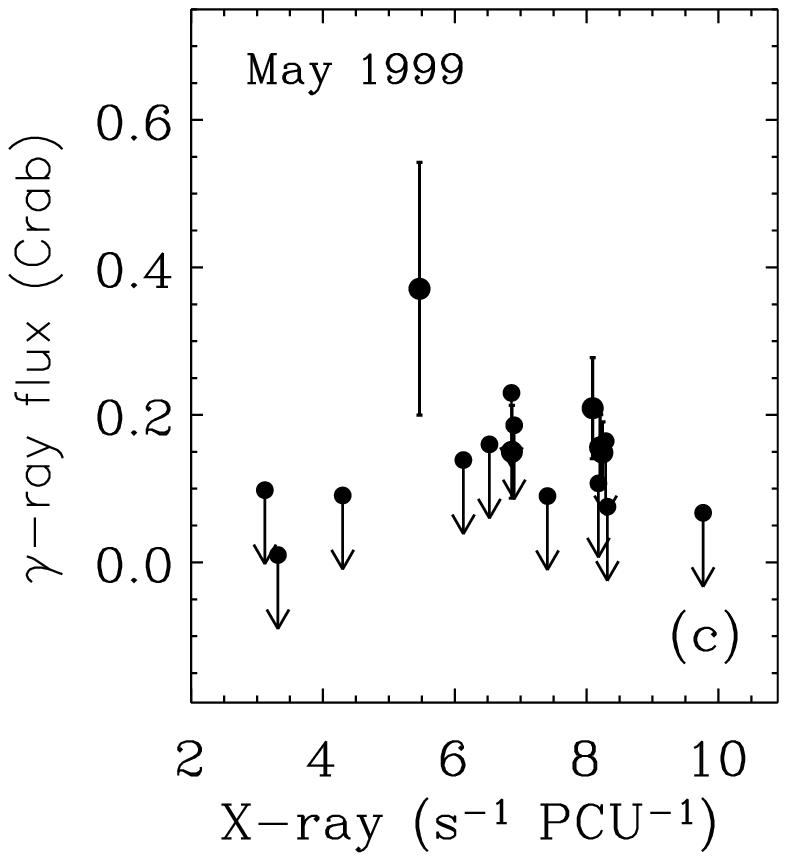}
\caption{Mkn\,501 TeV flux vs. X-ray count rate for three
representative flares, showing (a) a linear correlation
$F_{\gamma}\propto F_{\rm X}^{0.99\pm0.01}$, (b) a nonlinear
correlation $F_{\gamma}\propto F_{\rm X}^{2.07\pm0.36}$, and (c)
no correlation at all due to a rather low source flux and
predominantly upper flux limits only. Taken from
\cite{gliozzi05}.} \label{fig:9}
\end{figure}

\paragraph{Mkn 501 during 1997 -- 1999.}
Mkn 501 was studied over a period of 3 years from 1997 to 1999
\cite{quinn99,aharonian05}. Like for Mkn 421, a sampling every 3
to 4 days over longer observing periods was performed. For this
study, contemporaneous data not more than 8 hours apart, were
used. While a general positive trend was observed, a rather large
scatter (Fig.~\ref{fig:8}a) points to the presence of uncorrelated
activity; note also that one data point is $11\sigma$ off the
correlation (Fig.~\ref{fig:8}b). In this study also a general
correlation (Fig.~\ref{fig:8}c) over various observation periods
was established. When considering individual subperiods, a
heterogeneous behavior is found (Fig.~\ref{fig:9}): While in 1997
May a linear correlation is seen, $F_\gamma \sim F_{\mathrm
X}^{0.99\pm0.01}$, in 1998 June the correlation is quadratic,
$F_\gamma \sim F_{\mathrm X}^{2.07\pm0.36}$. Note that for 1999
May not enough data are available for claiming any correlation. In
fact, a time-dependent modeling of Mkn 501 spectra
\cite{krawczynski02} requires a two-zone SSC model, implying that
more complex correlations may be expected. Note also that in the
1999 observations a ``childless'' X-ray flare (without TeV
counterpart, that is), was found \cite{gliozzi05}.

\subsection{Recent Campaigns}
The unprecedented sensitivity of third-generation ground-based
instruments \cite{wystan}, allows for detailed studies of
correlations by a dense sampling both in the X-ray and VHE
bandpass. Particularly during flares, now light curves with a time
resolution of minutes are achieved. A multi-wavelength campaign on
the Southern source PKS 2155--304 during 2003 October 19 to
November 26 with strictly overlapping data sets from the High Energy
Stereoscopic System (H.E.S.S.) and {\em RXTE}-PCA revealed no
correlation \cite{aharonian05}, although intra-night variability
in the VHE band was found as well as an X-ray transient lasting
$\approx 1500$~s. A reason for the non-observation of a
correlation might be the rather small dynamical range of the
observations. In another multi-wavelength campaign in 2004
\cite{giebels07b}, a correlation between the $>200$~GeV and the
$2-10$~keV band was found with a correlation coefficient of
$r=0.71\pm0.05$.
\begin{figure}
\centering
\includegraphics[width=.55\linewidth]{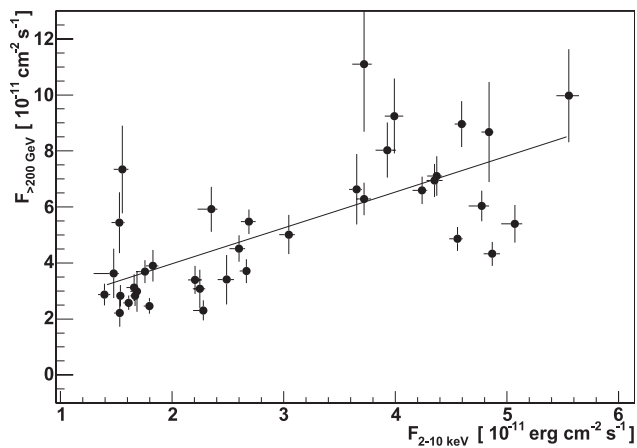}
\caption{VHE vs. 2--10 keV X-ray flux for PKS\,2155--304 in the
2004 multi-wavelength campaign during a $\gamma$-ray high state.
The correlation factor is $r=0.71\pm0.05$. Taken from
\cite{giebels07b}.} \label{fig:55}
\end{figure}
On 2006 July 28, PKS\,2155--304 underwent a dramatic flare,
reaching a peak level of about 15 times the Crab nebula flux
\cite{hess2155}. The overall flare complex that was recorded
lasted for about 60 minutes, with distinct sub-flares with
doubling times on the minute-scale. Two nights after the reported
flare, another outburst could be recorded, higher in peak 
intensity than the first flare, this time with simultaneous coverage in
the X-ray energy band by {\it Chandra} (the ``Chandra night'').
Among a wealth of interesting observational facts \cite{rolfdpg},
a correlation of the X-ray and VHE data was without any time lag.
The optical flux followed the VHE flux with a longer time
constant. The variabilities relate as $\Delta_\mathrm{VHE} :
\Delta_\mathrm{X} : \Delta_\mathrm{vis} \approx 14 : 2 : 0.15$.
Very surprisingly, an almost cubic correlation $F_\gamma \sim
F_{\mathrm X}^\alpha$, $\alpha\approx3$, was found during the
flare decay phase \cite{wystan}, which practically rules out a
simple one-zone SSC model for that observations.

In summer 2005, the Major Atmospheric Imaging Air Cerenkov (MAGIC)
telescope set out for an unbiased six-week fixed-window
observation of Mkn\,501 \cite{albert501}. The measured flux of Mkn
501 varied between $\approx 30\%$ and $\approx 4$~Crab units. As
Mkn\,501 was found on a moderate flux level, a day-by-day
correlation with the {\em RXTE}-ASM data revealed a marginal
correlation (dominated by the large errors of the ASM-measured
X-ray flux).

Also recent measurements of Mkn 421 confirm the existence of a
correlation: In 13 nights from November 2004 to April 2005, MAGIC
and ASM observed a correlation \cite{magic421}, as did, with the
mores sensitive X-ray instrument PCA, VERITAS during 14 hours from
2008 January 8 to February 13 \cite{lreyes08}. This data set provides
11 strictly overlapping X-ray/TeV data points, or, when relaxing
the simultaneity requirement by $\approx~5$~hours, 16 data points;
a correlation coefficient of $r=0.76$ is reached.

While for Mkn 421 and during high-flux episodes also for Mkn 501,
ASM X-ray data may be used for correlation studies, the ASM
instrument generally is not sensitive enough to allow such studies
for other sources (as demonstrated by \cite{magic2344}). For the
historically third-detected TeV-detected blazar, 1ES\,2344+514, only few
observational data exist, but a 3-month detection in a rather low
state was achieved by MAGIC in 2005 \cite{magic2344}. From October
2007 to January 2008 simultaneous VERITAS and {\em RXTE}-PCA
observations were conducted and resulted in a detection of a
simultaneous X-ray/$\gamma$-ray flare in that source; the
correlation coefficient for X-ray/$\gamma$-ray flux was found to
be $r=0.62$ \cite{horan08}. Interestingly, in the latter
observations an emission level quite compatible with the MAGIC
measurements was found.

The first non-blazar object detected in VHE $\gamma$-rays is the
close (16~Mpc) Fanaroff-Riley~I-type radio galaxy M\,87. This
object is believed to be a misaligned blazar \cite{tsv98}, showing
observational properties (e.g., short-timescale variability)
similar to blazars. For M\,87 the location of the VHE emission is
still uncertain: Both the region very close to the central engine,
and the brightest knot in the resolved jet, called HST-1, are
considered.
Note also that specific emission models for the misaligned-blazar
case have been developed (e.g., \cite{gpk05}. See \cite{albertm87}
for an overview). A $\gamma$-ray variability timescale as short as
1 day \cite{hessm87,albertm87} narrows down the size of the
emission region to be on the order of the light-crossing time of
the central blach hole, implying a production region in the
immediate vicinity of the M\,87 core. During observations
performed with a similar sensitivity, no significant flux
variations were found \cite{acciari08}. For M\,87, an X-ray -- VHE
$\gamma$-ray correlation is expected in most emission models, but
was not unambiguously found so far. Whereas \cite{hessm87} claim a
hint of a correlation between the soft (0.3-10 keV) X-rays at
HST-1 and the VHE $\gamma$-rays, \cite{acciari08} instead find a
year-by-year correlation between the (2$-$10 keV) X-ray flux of
the M\,87 core and the VHE $\gamma$-ray emission, but do not
observe a correlation between the two energy bands on shorter
timescales.
Interesting new data from a H.E.S.S.-MAGIC-VERITAS campaign in
early 2008 may strengthen the core-origin hypothesis of the
emission, as a high-level VHE flux and flares seen during this
campaign \cite{albertm87,hmvm87} were accompanied by a high-level
X-ray flux from the M\,87 core, while HST-1 was at its historical
low-level.

\section{Optical correlations, optical triggers, radio correlations}
No obvious correlations between the optical and the TeV regime
were found in 1ES~1959+650 \cite{krawczynski04}, Mkn 501
\cite{petry00,albert501}, PKS 2155--304 \cite{aharonian05}, and Mkn 421
\cite{giebels07}. It could, however, be shown that the fractional
root mean square variability amplitude obeys a power-law for Mkn
421 \cite{giebels07} and can thus be used as a helpful proxy. The
increase in fractional variability is also seen in X-rays
\cite{fossati00} and VHE $\gamma$-rays \cite{albert501}.

In recent years, however, quite some new TeV $\gamma$-ray emitters
have been detected following triggers indicating a high state of
these sources in the optical regime (Mkn 180 \cite{albert180}, 1ES
1011+496 \cite{albert1011}, S5~0716+714 \cite{magics5}; see
\cite{elina} for an overview). Additionally, the low-peaked blazar
BL~Lacert\ae\,\cite{magicbllac} and the FSRQ 3C~279
\cite{albert3c279} were successfully detected during comparatively
high states in the optical. It remains to be seen whether for some
of the recently-detected blazars, particularly for low-peaked
objects as 3C 279, BL~Lacert\ae, or W~Com\ae\ \cite{wcomae}
correlations with the optical regime can be established.

A connection of TeV activity and the blazar core variability
measured with Very-Large-Baseline Interferometry (VLBI) has
recently been established \cite{charlot06}. Contemporaneous
observations of Mkn 421 with the CAT instrument in the TeV energy
range and with VLBI in March-April 1998 resulted in no temporal
changes of the VLBI jet maps, but in strong evidence for total and
polarized flux variability of the VLBI core of Mkn 421 (on
timescales of weeks). The correlation is not yet established on a
firm and quantitative level, but the 22 GHz VLBI core seems to be
the self-absorbed counterpart of the SSC emission at higher
energies. Thus, VLBI can help to map the SSC zone. Lower radio
frequencies (15, 8, and 5 GHz) are found to be unrelated to the
SSC phenomenon, and originate likely outside the $\gamma$-ray
emission zone.

\section{Summary}
The up to now best-studied objects concerning correlations of the
two bumps in the spectral energy distribution of blazars all
belong to the class of high-peaked BL~Lac objects. These bright
TeV blazars have been studied in great detail for about the last ten
years on various timescales and during various emission-level
episodes.
Correlated variability up to now has been observed on timescales
ranging from hours to months, up to years and seems to be an
observationally established fact for the well-studied TeV blazars.
Most of the results, however, still suffer from various possible
experimental caveats, including the need for an
as-simultaneous-as-possible time coverage and a good sensitivity
of the instruments included in the observational campaigns. Last
but not least, one needs some luck to be able to follow a blazar
during an outburst and be able to infer the physics of the flare
production and decay.

There have been, however, few campaigns, as e.g. the 2001 campaign
on Mkn 421, which provided the possibility to study correlations
during the rise and decay phase of individual flares;
particularly also the observation of the 2006 flares of PKS
2155--304 seem to offer almost a dissecting knife for studying the
quantitative imprints of the physics processes that produce flares
on the correlations.

At the same time, interesting phenomena like ``orphan flares''
without X-ray counterparts and ``childless X-ray flares'' need to
be explained, and more generally, a systematic understanding of
lags between the two photon populations is still elusive.

While correlations of the emission on the high-energy tails of the
two photon populations have been studied in great detail, lately
also potential correlations of VHE $\gamma$ rays with the optical
or radio emission have been investigated. Optical triggers seem to
be a successful proxy to finding new TeV emitters, while first
studies of radio-TeV correlations seem to help locating the region
in which the SSC phenomenon takes place, profiting from the high
spatial resolution that VLBI can offer.

\acknowledgments I am thankful for helpful comments from Fabrizio
Tavecchio.

\end{document}